\documentclass[12pt,preprint]{aastex}
\usepackage{epsfig}
\usepackage{lscape}
\usepackage{rotating}



\lefthead{Nicastro et al.}

\begin{document}

\title{A Distant Echo of Milky Way Central Activity closes the Galaxy's Baryon Census}

\author{F. Nicastro$^{1,2}$, F. Senatore$^{[1,3]}$, Y. Krongold$^4$, S. Mathur$^{[5,6]}$, M. Elvis$^2$}

\altaffiltext{1}{Osservatorio Astronomico di Roma - INAF, Via di Frascati 33, 00040, Monte Porzio Catone, RM, Italy}
\altaffiltext{2}{Harvard-Smithsonian Center for Astrophysics, 60 Garden St., MS-04, Cambridge, MA 02138, USA}
\altaffiltext{3}{Università degli Studi di Roma “Tor Vergata”, Roma, Italy}
\altaffiltext{4}{Instituto de Astronomia, Universidad Nacional Autonoma de Mexico, Mexico City, Mexico}
\altaffiltext{5}{Ohio State University, Columbus, OH, USA}
\altaffiltext{6}{Center for Cosmology and Astro-Particle Physics, Ohio State University, Columbus, OH 43210, USA}

\begin{abstract}
We report on the presence of large amounts of million-degree gas in the Milky Way's interstellar and circum-galactic medium. 
This gas (1) permeates both the Galactic plane and the halo, (2) extends to distances larger than 60-200 kpc from the center, and 
(3) its mass is sufficient to close the Galaxy's baryon census. 

Moreover, we show that a vast, $\sim 6$ kpc radius,  spherically-symmetric central region of the Milky Way above and below the 0.16 kpc thick plane, has either 
been emptied of hot gas or the density of this gas within the cavity has a peculiar profile, increasing from the center up to a radius of $\sim 6$ kpc, and then decreasing 
with a typical halo density profile. This, and several other converging pieces of evidence, suggest that the current surface of the cavity, at 6 kpc from the Galaxy's center, 
traces the distant echo of a period of strong nuclear activity of our super-massive black-hole, occurred about 6 Myrs ago. 
 \end{abstract}

\keywords{}

\section{Introduction}
The visible baryonic mass of the Milky Way, which includes stars, cold and mildly photo-ionized gas and dust, amounts to 
M$_b^{Obs} \simeq 0.65×\times 10^{11}$  M$_{\odot}$ (McMillian \& Binney, 2012). 
The total baryonic plus dark matter mass of our Galaxy, is instead estimated in the range M$_{Tot} \simeq (1-2) \times 10^{12}$  M$_{\odot}$  (Boylan-Kolchin et al., 2013). 
This, assuming a universal baryon fraction of $f_b = 0.157$ (The Plank Collaboration, 2015), implies a total baryonic mass of M$_b^{Pred} \simeq (1.6-3.2) \times 10^{11}$  
M$_{\odot}$, between 2.5 and 5 times larger than observed. A large fraction of the baryonic mass of our Galaxy seems to be currently eluding detection. 

This “missing-baryon” problem is not a monopoly of the Milky Way: most of the galaxies in the local universe suffer a deficit of baryonic mass compared to their 
dynamical mass and the problem is more serious at smaller dynamical masses (e.g. McGaugh et al., 2010), suggesting that lighter galaxies fail to retain larger fractions 
of their baryons. 
Recurring activity during the galaxy's lifetime, such as bursts of star formation followed by powerful supernova explosions or accretion-powered ignitions of the 
central supermassive black hole, may power energetic outflows that heat the gas to high temperatures ($\sim 10^6$ K) and push out material from the galaxy's 
metal-rich innermost and star-populated regions to its surrounding volume, perhaps up or beyond the galaxy's virial radius. 
Evidence of such warm-hot and metal-enriched baryons is strong, both in the vicinity of external galaxies (Lehner et al., 2013; Stocke et al., 2006; Prochaska et al., 2011), 
where Far-Ultraviolet (FUV) observations are sensitive to metals with temperatures in the T$\simeq (5-500) \times 10^3$ K range, and in our own Milky Way (Nicastro et al., 
2002; Williams et al., 2005; Bregman \& Lloyd-Davies, 2007; Gupta et al., 2012; Miller \& Bregman, 2013; Fang et al., 2013, 1015; Yao \& Wang, 2007; Faerman, Sternberg 
\& McKee, 2015), where X-ray observations trace hot metals with temperatures in the T$\simeq (0.5-5) \times 10^6$ K range.  

In particular, over the past several years, a number of experiments, as well as theoretical works, have attempted to gain insights into the location and mass of the 
hot medium in our own Galaxy. This is not a trivial task, not only because the available observables (namely, OVII column densities - e.g. Bregman \& Lloyd-Davies, 2007 -, 
OVII Emission Measure – e.g. Gupta et al., 2012 - and the Pulsar Dispersion Measure – e.g. Fang et al., 2013) are all degenerate in path-length crossed through the medium 
along our line of sight and density of the medium itself, but also because the density distribution of the medium is unknown; certainly assuming it to be constant throughout 
the whole Galaxy and out to its virial radius (e.g. Nicastro et al., 2002; Williams et al., 2005; Gupta et al., 2012) is not physically justified. 
However, our peripheral position in the Galaxy, at about 8.5 kpc from the Galaxy's center and roughly in the Galaxy's plane, gives us hope of solving the problem: once 
a physically motivated density profile is assumed for the hot absorbing medium, the observed column densities (as well as the other observables) will depend critically 
on the sky position (and distance, for Galactic background targets) of the sources towards which the column densities are measured. For example, any spherically or 
cylindrically symmetric (with respect to the Galaxy's center) density profile would imprint stronger absorption, as seen by us, in the direction of the Galaxy's center than 
towards the anti-center. This consideration has recently motivated several studies, which have used available spectra of extragalactic targets, with no other selection 
criterion than being at high Galactic latitudes, to measure OVII column densities and compare them with physically motivated or simple phenomenological density profile 
models (Miller \& Bregman, 2013; Fang et al., 2013; Faerman, Sternberg \& McKee, 2015). 
The results, however, are often contradictory, with estimated total masses of the million degree medium within a 1.2 virial-radius sphere (300 kpc) 
that strongly depend on the flatness of the assumed density profile, and range from a negligible M$_{Hot} \simeq 2.4 \times 10^9$  M$_{\odot}$ (Miller \& Bregman, 2013) to 
a significant M$_{Hot} \simeq 10^{11}$ M$_{\odot}$ (Faerman, Sternberg \& McKee, 2015).  

\noindent
We argue that such large differences are mostly due to the impossibility, when using only observables towards high Galactic latitude lines of sight, of anchoring the 
value of the baryon density of hot material in the central region of the Galaxy to its actual value in the Galaxy's plane. 
For this reason, here we perform an experiment similar to that of Miller \& Bregman (2013), but for the first time simultaneously fitting OVII absorption towards both high-latitude 
(HGL sample), and low-latitude (LGL sample) lines of sight, respectively against background Active Galactic Nuclei (AGNs) and Galactic X-Ray Binaries (XRBs) with known 
distances (Figure 1). 

\noindent
Throughout the paper, we refer to all densities and masses in units of ($A_O/4.9\times 10^{-4})^{-1} \times [Z/(0.3 Z_{\odot})]^{-1} (f_{OVII}/0.5)^{-1}$, where $A_O$ is the relative abundance 
of oxygen compared to hydrogen, $Z$ is the metallicity and $f_{OVII}$ is the fraction of OVII relative to oxygen. 
For easy comparison to other works (e.g. Faerman, Sternberg \& McKee, 2015), we compute hot baryon masses within a 1.2 virial-radius sphere. 
Errors on best-fitting parameters (and quantities derived from those) are provided at 90\% confidence level for a number of interesting parameters equal to (31-N$_{dof}$), where 
N$_{dof}$ is the number of degrees of freedom in the fit.

\section{Data, Observables and Modeling} 
Here we briefly outline the data, the observables and the specific procedures we use in our analysis, and refer instead to a forthcoming paper on the analysis of all the warm-hot 
components in the disk and halo of our Galaxy (Senatore et al., 2016a; in preparation), for a detailed description of the full data-set, its reduction 
and analysis. 

\subsection{Sample Selection and OVII K$\alpha$ Absorbers}
To perform our analysis, we mined the XMM-{\em Newton} Science archive. This led to select two signal-to-noise limited samples of Reflection Grating 
Spectrometer (RGS) spectra: one of low Galactic latitude XRBs (LGL sample) and one of high Galactic Latitude AGNs (HGL sample).  

Our total sample differs from those previously used to perform analyses similar to ours (e.g. Gupta et al., 2012; Miller \& Bregman, 2013) in three important ways: 
(1) for the first time we use simultaneously HGL and LGL samples. 
(2) our two RGS spectral samples are complete to a minimum Signal to Noise per Resolution Element in the continuum, SNRE$>$10 at 22 \AA\ (just longward of 
the local OVII K$\alpha$): this makes our spectra sensitive to unresolved line Equivalent Width EW$>20$ m\AA\ at 3$\sigma$, and thus allows the 
detections of the K$\alpha$ and K$\beta$ transitions of OVII in absorption (our main observables) at relatively high statistical significance. 
Previous works (e.g. Gupta et al., 2012; Miller \& Bregman, 2013) considered spectra with EW sensitivity even one order of magnitude larger than 
ours, which led to severely over-estimated line EWs, and so column densities, in the poorest SNRE spectra (Senatore et al., 2016b; in preparation). 
(3) whenever possible (see below), we  remove the degeneracy between column density and Doppler parameter of the instrumentally unresolved 
OVII lines, by performing a detailed curve of growth analysis (e.g. Nicastro et al., 2016). 

Requiring SNRE$>10$ led to a total of 51 HGL and 20 LGL targets suitable for analysis. Of the 20 LGL spectra, 18 (90\%) show the presence of local OVII Kα absorption. 
Of these 18 LGL targets, only 14 have known distance, and one of these 14 (PSRB~0833-45) is too nearby (300 pc) for the observed amount of OVII absorption be entirely 
produced in the intervening ISM. We therefore conservatively exclude PSRB~0833-45 from our LGL sample, which is thus made of 13 lines of sight (Figure 1, filled red circles). 
Of these, 11 have OVII K$\alpha$ detected with significance $>3\sigma$; the other 2 have OVII K$\alpha$ at significance $>2\sigma$. 
Of the 51 HGL targets, 9 are low-redshift and therefore the presence of an intrinsic warm-absorber contaminates the local OVII absorption in their spectra. 
Three of the remaining 42 targets have RGS spectra affected by an instrumental feature at the rest frame position of the OVII K$\alpha$ transition, and so are excluded from 
the analysis. 
Of the remaining 39 HGLs, 34 (87\%) show OVII K$\alpha$ absorption. However in 8 of these the measured OVII K$\alpha$ equivalent-width (EW) has statistical significance 
$<2\sigma$. Given the large number of lines of sight available for the HGL sample, compared to that of LGL sample, we exclude these targets from our analysis. Finally, only 14 of 
the remaining 26 spectra show both the K$\alpha$ and K$\beta$ lines, but 4 of the 12 that do not show the K$\beta$, have the K$\alpha$ detected at $>3\sigma$. 
We use these 14+4 HGL spectra in our analysis (Figure 1, filled blue circles). 
Hence our final HGL and LGL samples contain 18 and 13 lines of sight, respectively, leading to a well-defined SNRE-complete observed distribution of 31 OVII K$\alpha$ 
EWs and sky positions. 

\subsection{Converting OVII K$\alpha$ EWs to Column Density}
The local OVII bound-bound absorbers are all unresolved at the RGS resolution (FWHM = 950 km s$^{-1}$ at 22 \AA). Consequently, for all saturated OVII lines, the line of 
sight column density N$_{OVII}$ is degenerate with the Doppler parameter $b$. Removing this degeneracy, to properly convert the measured EW into a reliable column density, 
is crucial for this kind of analysis. The optimal way would be to evaluate $b$ and N$_{OVII}$, for each line of sight independently. This is possible whenever at least two transitions 
from the same ion are detected, and their combined statistical significance is sufficient to provide a solution (e.g. Nicastro et al., 2016). 
In our samples, we are able to properly evaluate $b$ and N$_{OVII}$ independently for 13/18 HGL and 9/13 LGL lines of sight. For the remaining 9 (5 HGLs and 4 LGLs), 
we evaluate OVII column densities by weighting the average of the Doppler parameters measured for our HGL and LGL samples, over the combined statistical significance 
of the OVII K$\alpha$ and K$\beta$  EWs: $<b_{HGL}> = 95$ km s$^{-1}$ and $<b_{LGL}>= 125$ km s$^{-1}$. 
We note that our HGL averaged $b$ value is similar to the average $b$ value found by Fang et al. (2013) and assumed by Faerman, Sternberg \& McKee (2017), but rather different 
from the large Doppler parameter of 150 km s$^{-1}$ assumed by Miller \& Bregman (2013) for all the sources of their sample. 
A large value of $b$, especially if applied to all the lines of sight, not only introduces a severe bias 
towards small column densities, but may also artificially modify the column density versus sky-position distribution, which is crucial to constrain the density profile of 
the absorber (see \S 2.3-2.5). 

\noindent
Our HLG OVII absorbers spread over more than an order of magnitude in column densities, from a minimum value of N$_{OVII} = 0.8_{-0.5}^{+1.2} \times 10^{16}$ cm$^{-2}$, to a 
maximum value of N$_{OVII} = 33_{-29}^{+480} \times 10^{16}$ cm$^{-2}$. 
The spread is less extreme for LGL absorbers that span a factor of about 5 in OVII column densities, with the 
exception of one outliner, the low-mass XRB V*V821~Ara, for which, however, contamination by an intrinsic absorber has been proposed (Miller et al., 2004). 

\subsection{Functions and Data-Matching Procedure}
To model our data we do not rely on any particular fitting optimization methods and use a statistical approach that allows us to perform a full posterior 
analysis of the probability function. 
For each data-matching run, indeed, we build the entire M-dimensional (where M is the number of parameters of the given functional form), probability function 
(Likelihood) $f(p_1,...p_M) \propto e^{-\sum_{k}^{} \Delta \chi^2_k}  = e^{-[N_k^{Obs} - N_k^{Mod}(p_1,...,p_M)] / \sigma_k^2}$ (where $N_k^{Obs}$ and $N_k^{Mod}$ 
are the observed and model-predicted OVII column densities at a given sky position, and the sum extends over the number of sky positions used in the run: i.e. 
13 for LGL, 18 for HGL and 31 for LGL+HGL runs) over the full M-dimensional parameter space. 
Each of the M model parameters are varied within wide uniform (i.e. equiprobable priors) intervals whose boundaries are set by astrophysical priors (e.g. 
the lower and upper boundaries of the core radius interval are set to 10 pc and the Galaxy's virial radius, respectively). 
At each iteration (i.e. for a given set of model parameter values and for each line of sight direction) within a single run, $N_k^{Mod}$ are evaluated by integrating 
our functional form density profile $n(R)$ along the line of sight coordinate $\xi(R)$ [with $R^2 = \xi^2 + R_{\odot}^2 - 2\xi R_{\odot}  cos⁡(b)cos⁡(l)$]. 
For each data-matching run, we therefore dispose of the full M-dimensional Likelihood function (with size M$\times$L$_1 \times$...L$_M$, where L$_j$ is the 
number of equiprobable values of the $j^{th}$ parameter), which we use: (1) to extract the set of paramater values that maximizes the Likelihood (or minimizes 
$\sum_{k}^{} \Delta \chi^2_k$: i.e. to define the best-matching model), (2) to estimate parameter confidence intervals and (3) to investigate on their degeneracies 
and necessity.
 
We make use of two most general families of density profiles, for material embedded in deep gravitational potential wells (like the dark-matter potential well of our 
Galaxy), i.e.: spherically-symmetric (SS), where the only scale-length parameter is the core-radius $R_c$, and Cylindrically-Symmetric (CS), characterized by two different 
scale-length parameters, the coplanar core-radius $\rho_c$ and the vertical core-height $h_c$. 
In particular, we use the following four different and extremely general phenomenological functional forms for our density-profile models: 
(1) a SS exponential profile, characterized by two parameters: the normalizing density $n_0$ (in cm$^{-3}$) and the core radius $R_c$ (in kpc): 
$$n(R) = n_0  e^{-|R-R_s |/R_c} ;$$
(2) a flattened CS exponential profile, characterized by three parameters: the normalizing density $n_0$, the coplanar core radius $\rho_c$ (in kpc) and the 
vertical core-height $h_c$ (in kpc): 
$$n(R) = n_0  e^{-\sqrt{ (\rho/\rho_c )^2 + (z-h_s)/h_c )^2}} ;$$
(3) a SS β-profile, with the three parameters $n_0$, $R_c$ and the index $\beta$ characterizing the steepness of the profile: 
$$n(R) = n_0 [1+(R-R_s )^2/R_c^2 ]^{-3\beta/2} ;$$
(4) a flattened CS β-profile, with the four parameters $n_0$), $\rho_c$, $h_c$ and $\beta$: 
$$n(R)= n_0 [1+\rho^2/\rho_c^2 + (z-h_s )^2/h_c^2 ]^{-3\beta/2} .$$

As shown in the analytical expressions, for each of these four functional forms we also allow for the inclusion of an additional parameter ($R_s$ for the SS profiles and 
$h_s$ for the CS profiles, both in kpc) allowing for a possible offset of the distributions from the Galaxy's center (SS models) or plane (CS models).  

Through their parameters $\beta$, $R_s$ and $h_s$ (and at a less extent the core distances $R_c$, $\rho_c$ and $h_c$) these functional forms describe a wide spectrum of 
radial density trends (from flattened exponential disks, to steep isothermal halos - $\beta =2/3$ -, to flat hydrostatic equilibrium gas distributions - $\beta \simeq 1/4$) and 
structures (peaks and valleys). 


\subsection{Determination of Halo Extent and Masses}
Given the one-dimensional nature of our observables (and, for the HGLs, the a-priori unknown maximum line of sight integration distance), and so the need for reducing 
the potentially 3-dimensional models to the single line of sight direction during the fitting procedure, only a lower limit to the total extent of the volume containing 
the hot absorbing gas seen against HGL targets can be evaluated in our analyses. We evaluate this limit by stopping the line of sight integration at a line of sight distance 
$\xi$ where the relative difference between two consecutive values of the column density differs by less than 0.01\% (much lower than the typical relative uncertainty on our 
column density measurements, which is of the order of $\simeq 10$\% in the best cases). Under the assumption of a centrally symmetric halo, the largest of these line of 
sight distances in the best-fitting HGL models, sets effectively a lower limit to the maximum radial size of the halo, and so its baryon mass. Smaller halos are not allowed 
by the necessity to accumulate sufficient column density (and emission measure) along the thickest HGL lines of sight. On the other hand, larger halo sizes (and therefore 
baryon masses) are clearly possible, but not directly measurable through our observables. 

\subsection{Parameter Degeneracy}
For the same limitation intrinsic in the one-dimensionality of our observables, the parameters of our models are all degenerate, to some extent. 
For exponential profiles, where the flatness of the distribution is not parameterized by a varying parameter, the problem is negligible and only a moderate degeneracy 
is present between the peak density $n_0$ and the scale-distance parameters $R_c$ or $\rho_c$ and $h_c$ (such degeneracy is practically absent for LGLs, where the 
distance of the background targets is known). 
For $\beta$-like profiles, that are characterized by one degree of freedom less compared to their corresponding exponential profiles, the scale-distance parameters are 
often strongly degenerate with the flatness index $\beta$ and, when this happens (typically in our HGL-only runs - \S 3.1 - or in the combined LGL+HGL runs with offset radius 
forced to be zero - \S 3.2), the $\chi^2(R_c,\beta)$ surface does not display an absolute minimum but rather an asymptotic minimum for diverging $R_c$ and $\beta$ values, 
and is impossible to discriminate statistically between very steep and compact (exponential) or flat and extended profiles. In these cases exponential profiles or $\beta$-like 
profiles with either the flatness index $\beta$ or radial scale distance $R_c$ frozen to some physically motivated value, provides the only non-degenerate solutions 
(i.e. model M1 in Table 1 or models M3 and M4 in Table 2). 
In all cases, however, the simultaneous modeling of low and high Galactic latitude lines of sight, together with presence of a radial offset in the distribution (i.e. best-fitting 
$R_s > 0$), tend to break the degeneracy between scale-distance and the index $\beta$. When this happens, the shape of the density profile can be determined and all model 
parameters are generally robustly constrained to physically reasonable best-fitting values (models A and B in table 2), and so are the derived minimum extent and mass of 
the halo. 

\section{Results}
\subsection{Separate Disk and Halo Fitting}
First we modeled the HGL and LGL separately. The 18 HGL absorbers are equally well fitted by both SS and CS models, and the best-fitting core radius and height, in CS 
models, are fully consistent with each other within the uncertainties, indicating that a flattened distribution is not statistically required. Density profiles are generally steep 
and, interestingly, the best-fitting profile is exponential and has an offset-radius $R_s =5.4_{-0.4}^{+0.6} $ kpc (Table 1, model M1). This shift in radius would indicate that the 
hot baryon density in the halo increases radially from the Galaxy's center up to its peak value at 5.4 kpc, and then decreases monotonically towards the virial radius. 
The implied baryonic mass is unimportant $M_{Hot}^{Halo} =3.3_{-1.4}^{+4.1}  \times 10^9$ M$_{\odot}$ (in full agreement with the mass derivable from the best-fitting 
parameters of the “spherical-saturated” model of Miller \& Bregman, 2013: see their table 2). 

Unlike the HGL absorbers, for the 13 members of our LGL sample a flattened disk-like CS density profile is statistically greatly preferred to a SS profile [$\chi_{Flat}^2 = 12.8$ 
for 10 degrees of freedom (dof), versus $\chi_{Sph}^2(dof) = 22.5(11)$]. The LGL absorbers are clearly tracing a disk-like distribution in the Galaxy's plane, with best-fitting 
radial and height scale lengths (Table 1, Model M2) in excellent agreement with those of the stellar disk of the Milky Way  (Rix \& Bovy, 2013). 
The mass of the hot gas in the Galactic disk is only $M_{Hot}^{Disk} = 1.4_{-0.6}^{+1.1}  \times 10^8$ M$_{\odot}$, of the order of that of the other gaseous components of the 
disk (Rix \& Bovy, 2013). 

The models that best-fit separately our HGL (halo) and LGL (disk) absorbers, are very different in both their central densities and profiles (Tab. 1), and neither of the two can 
adequately model the column-density distribution of the other. Simple extrapolation of the best-fitting disk model M1 to the HGL absorbers gives $\chi^2(dof) = 373.2(18)$ 
while the opposite gives $\chi^2(dof) = 147.0(13$). 
Either a compromise single-component model is needed, or the two components must be physically distinct

\subsection{Simultaneous Disk and Halo Fitting}
We then proceeded to model simultaneously and self-consistently all 31 HGL and LGL lines of sight. 
We tried two alternative families of functions: (a) a single-component set of models, with all parameters free to vary and the normalizing density peaking at 
$R = R_s$ (A-type models, hereinafter), and (b) a 2-component set of models in which a parameter-varying SS component with $n=0$ for $R<R_s$ and 
$n = n(R)$ for $R \ge R_s$, is added to a flattened disk-component with parameters frozen to the LGL best-fitting values (B-type models, hereinafter). 
Both sets of models can provide statistically acceptable fits (A and B in Table 2). In both cases offset radii $R_s > 0$ are statistically preferred 
(compare A with M4 and B with M3 in Table 2). 

\subsection{Statistical Significance of $R_s > 0$ and the Need for its Presence}
Our two best-fitting models A and B have offset radii $R_s =5.6_{-0.6}^{+0.6}$  kpc and $R_s = 6.7_{-1.8}^{+0.9}$ kpc (Table 2), consistent with each other and 
with the best-fitting value found by fitting the HGL sample only (Table 1). 
Figure 2 shows that $R_s =0$ is ruled out at a 4-interesting-parameter statistical significance of 14.9$\sigma$. Similarly, for our alternative best-fitting model B, 
$R_s = 0$ is excluded at a 4-interesting-parameter statistical significance of 6.0$\sigma$. 
To evaluate the statistical “need” for an offset radius $R_s > 0$, we perform a comparison between our simpler models M3 and M4, with our complex best-fitting 
model A. Both models M3 and M4 are nested in model A. From Table 2, $\Delta\chi^2(A;M4) = 16.9$ and $\Delta\chi^2(A;M3) = 5.7$. These increments are for 2 
additional parameters and for a number of degrees of freedom of the complex model A of 27. The F-test probabilities that models M4 and M3 are to be rejected compared 
to model A, are therefore P(A;M4) = 99.8\% and P(A;M3)=91.3\%, respectively. 

We can also investigate on the the need for an offset radius $R_s>0$, from the point of view of the observables (the posteriors in a non-{\em frequentist} 
approach), through a comparison analysis with the full M-dimensional Likelihood function. The need for an off-set radius comes from the need of both reproducing 
simultaneously the LGL+HGL column density versus sky position distribution, and matching the large (factor of ≈40) spread observed in the column density distribution of 
HGL OVII absorbers between center and anti-center directions. Such a large spread cannot easily be reproduced with models in which the halo density peaks at the Galaxy's 
center, but is instead naturally reproduced by introducing a radial structure in the density profile that adds up to the geometrical structure induced by our peripheral position 
in the Galaxy to further modulate the column density versus sky-position dependence, and amplify the column density towards all those directions that go through two 
distinct density peaks. 
Our single-component model A is the one that best reproduces the observed $\Delta N_{OVII}(l,b)$ spread in our HGL absorbers, while still modeling sufficiently well 
the LGL columns. The corresponding $R_s = 0$ model M4, instead, reproduces less well the observed $\Delta N_{OVII}(l,b)$ spread in HGL absorbers 
and tend to systematically under-predict the column density of LGL absorbers (generally characterized by smaller uncertainties, compared to HGL absorbers). 
Conversely, our 2-component models B and M3, by construction, model well the LGL column densities 
but fail in reproducing the $\Delta N_{OVII}(l,b)$ spread of the HGL absorbers: in particular model M3 severely under-predicts most of the observed HGL column densities 
towards the Galactic center. 

\subsection{Best Fitting Models A versus B}
Model A is able to better reproduce the observed spread of OVII column densities along HGL lines of sight, compared to model B, which however (by construction) reproduces 
better the observed LGL columns. The actual solution lies probably in between models A and B. However looking for such compromising solution implies leaving the three 
parameters of the flattened exponential disk component free to vary simultaneously with those of the halo component, resulting in a minimum of 7 varying parameters. 
This makes the model too complex relatively to the quality of the data that we dispose, and its parameters too degenerate. The result is that the central density of the disk 
component is pegged to zero during the data-matching procedure, and the solution reduces to the single-component best-fitting model A. 
A Bayesian approach to this problem, would probably allow us to better quantify the ``degree of believes'' to which one of the two models is preferred to the 
other, but this is beyond the scope of this paper.

\subsection{Emission Measures}
Our best-fitting models A and, to a lesser extent, B also reproduce well the Emission Measure at high Galactic latitudes and towards the Galactic center. In a 1-steradian wide 
region of the sky around $l=90^0$, $b=60^0$, McCammon and collaborators (2002) measure EM = 0.0125 cm$^{-6}$ pc, while our models predict 
EM(A;$|b|\ge 60^0;30^0<l<150^0$)$ =0.013_{-0.009}^{+0.022}$ cm$^{-6}$ pc and EM(B;$|b|\ge 60^0;30^0<l<150^0$)$ = 0.005_{-0.003}^{+0.012}$ cm$^{-6}$ pc. 
Spatially resolved, absorbed (i.e. non local) X-ray emission is also clearly seen in a vast region around the Galactic center (e.g. Kataoka et al., 2015 and references therein; 
see also \S 4), with EM spanning values between (0.08-0.3) cm$^{-6}$ pc at $|b|<20^0$ (Kataoka et al., 2015). 
In the same region our models A and B predict  EM(A;$|b|\le 20^0;340^0<l<20^0$)$ =0.4_{-0.2}^{+0.1}$ cm$^{-6}$ pc and EM(B;$|b|\le 20^0;340^0<l<20^0$)$ =0.03_{-0.01}^{+0.03}$ 
cm$^{-6}$ pc, consistent, within the uncertainties, with the upper and lower boundaries of the observed interval, respectively, suggesting again that the actual solution lies 
somewhere in between models A and B. 

\subsection{Measurement of Halo Extent and Mass}
From our best-fitting models we derive total hot baryon masses in the ranges M$_{Hot}(A) =0.2_{-0.1}^{+0.3}  \times 10^{11}$ M$_{\odot}$ and M$_{Hot}(B) = 1.3_{-0.7}^{+2.1}  \times 
10^{11}$ M$_{\odot}$ (Table 2). These masses are $>10$ times larger than those obtained by fitting the HGL sample only (Table 1). This is due to the flatness of the best-fitting 
density profiles (Table 2): in model A, $\beta \simeq  2/3$, which implies $n(R) \simeq R^{-2}$, as in a simple isothermal sphere, while model B has an even flatter profile, 
with $\beta \simeq 1/3$, i.e. similar to those obtained by imposing hydrostatic equilibrium of the hot gas in the Galaxy (e.g. Fang et al., 2013; Faerman, Sternberg \& McKee, 2015). 
Such flat profiles results essentially from the need to simultaneously model column densities at low (LGL) and high (HGL) Galactic latitudes. 
This strong constraint plays slightly differently in a- and B-types models: in A-type models a single component must reproduce both LGL and HGL column densities. This implies 
that, in the innermost Galaxy's region, both the density and the scale distance must be sufficiently large to reproduce the observed LGL columns, but not too large to over-predict 
them. This leaves little room to model the typically larger HGL columns in a small central volume, and requires flat profile to extend the integrations at large distances 
(typically $>60$ kpc, Table 2). 
In B-type models, instead, both a disk and a halo component are present, with the disk component frozen to the best-fitting LGL model M2. The halo 
component must therefore be characterized by low normalizing densities (not to over-predict the observed LGL columns) and therefore can only reproduce the observed HGL 
columns by extending over a large volume, with radius of at least 200 kpc (Table 2). In both cases, a shifted density peak at $R_ss>0$, on one hand helps recovering the observed 
HGL columns without over-predicting the LGL ones and, on the other contributes to further flatten the profile to allow for the accumulation of large portions of the observed 
HGL column densities in large volumes outside the internal hollowed out sphere.

Adding the hot baryon mass to the visible mass of the Milky Way, gives a total baryonic mass in the range M$_b = (0.8-4.0) \times 10^{11}$ M$_{\odot}$, 
sufficient to close the Galaxy's baryon census (Fig. 3).

\section{Discussion and Conclusions}
Our analysis indicates not only (1) that both the Galactic plane and the halo are permeated by OVII-traced million degree gas, but also (2) that the amount of OVII-bearing 
gas in the halo is sufficient to close the Galaxy's baryon census and (3) that a vast, $\sim 6$ kpc radius,  spherically-symmetric central region of the Milky Way above and 
below the 0.16 kpc thick plane, has either been emptied of hot gas (Model B) or the density of this gas within the cavity has a peculiar profile, increasing from the center up 
to a radius of $\sim 6$ kpc, and then decreasing with a typical halo density profile (Model A). 

The large value of $R_s$ in both the scenarios implied by Model A and Model B can be understood in terms of a radially expanding blast-wave or a shock-front generated in 
the center of the Galaxy and traveling outwards, so acting as a piston onto the ambient gas, and compressing the material at its passage, while pushing it (or a fraction of it) 
outwards. The central black hole of our Galaxy, could have played a fundamental role in this (e.g. Dav\'e, Oppenheimer \& Finlator, 2011; Faucher-Gigu\'ere \& Quataert, 2012; 
Lapi, Cavaliere \& Menci, 2005), during a recent period of its activity. Faucher-Giguére \& Quataert (2012) study the property of galactic winds driven by active galactic nuclei, 
and show that energy-conserving outflows with initial velocity $v_{in} > 10000$ km s$^{-1}$, can move in 
the ambient medium producing shocked wind bubbles that expand at velocities of $v_s \simeq 1000$ km s$^{-1}$ into the host galaxy. 
If the observed OVII-bearing bubble in our Galaxy is tracing one of such shocks generated by our central supermassive black hole during a period of strong activity then, 
at a speed of 1000 km s$^{-1}$, the expanding shell would have taken 6 Myrs to reach its current radius of 6 kpc. Interestingly, $(6 \pm 2)$ Myr is also the age estimated for the 
two disks of young stars present in the central parsec of our Galaxy that are thought to be a relic of a gaseous accretion disk that provided fuel for AGN-like activity of our 
central black hole about 6 Myr ago (Paumard et al., 2006; Levin \& Beloborodov, 2003). 

Approaching to the problem from the opposite side, from simple considerations based on the energetic of such AGN outflows and the feedback between these and the 
surrounding ambient medium (e.g. Lapi, Cavaliere \& Menci, 2005), we can estimate the average velocity of the blast wave. 
According to Lapi, Cavaliere \& Menci (2005), and for a simple isothermal sphere, the mass contrast 
between the gas contained in the bubble and the total hot gas mass within one virial radius, equals about half of the energy contrast $\Delta E/E$ between the kinetic energy 
in the outflow (provided by the central AGN) and the total energy residing in the equilibrium hot gas. From our best-fitting model A (whose density profile decreases as in 
a simple isothermal sphere) the amount of gas currently contained in the bubble is M$_{Bubble} = 3.4_{-1.3}^{+1.8}  \times 10^8$ M$_{\odot}$. 
The contrast between this mass and the hot baryon masses filling a virial-radius sphere is then in the range $(\Delta M/M)_{Bubble} \simeq 0.005-0.05$. 
Thus the energy contrast is $\Delta E/E \simeq 0.01-0.1$. 
The total energy in the equilibrium hot gas is $E \simeq 2 \times 10^{61} ((kT_{Hot})/keV )^{5/2}$ ergs (Lapi, Cavaliere \& Menci, 2005), where $T_{Hot}$ is the temperature of the 
gas, which we assume in the range $T_{Hot} \simeq (0.6-1) \times 10^6$ K (where the OVII fraction peaks in gas in collisional equilibrium). 
Thus, $E \simeq (0.1-0.4) \times 10^{59}$ ergs, and $\Delta E \simeq (0.1-4) \times 10^{57}$ ergs $\simeq 1/2 \Delta M \overline{v_s^2}$ which gives $\overline{v_s} \simeq 
150-1400$ km s$^{-1}$, in good agreement with the $v_s = 1000$ km s$^{-1}$ needed to take the shock front from the Galaxy center to its current position of 
$R_s \simeq 6$ kpc, in 6 Myr. 
We also note that the above estimate of the kinetic energy of the outflow, $\Delta E \simeq (0.1-4) \times 10^{57}$ ergs, is significantly lower than (and so fully consistent 
with) the maximum amount of energy that can be made available from our Galactic nuclear black hole (M$_{BH} = 4.3 \times 10^6$ M$_{\odot}$ (Gillessen et al., 2009) by accreting 
at its Eddington limit for a period of 4-8 Myrs (limit imposed by the age of the two central star disks, under the assumption that they are relics of the gaseous accretion disk): 
$E_{AGN} (\Delta t=4-8 Myrs) \simeq (2.5-5) \times 10^{58}$ ergs. 

The OVII bubble mass can also be used to evaluate the outflow rate needed for the AGN to deploy such an amount of mass at 6 kpc from the nucleus. Again, assuming a 4-8 
Myrs long period of nuclear activity, this rate is $\dot{M}_{Out} \simeq 26-130$ M$_{\odot}$  yr$^{-1}$, only a factor of few lower than those typically observed in similar 
scale molecular outflows in external galaxies hosting nuclear black holes that are generally more massive than ours (27). 

The central spherical region traced by the hot OVII gas in the Galaxy, has both size and shape similar to those of two other peculiar structures recently discovered in microwave 
and Gamma-ray excess emission by the Planck and Fermi satellites: the so called Plank and Fermi Bubbles (Dobler et al., 2010; Su et al., 2010). 
These structures are co-spatial (extending for 4-10 kpc above and below the Galactic plane) and thought to be produced by the same population of hard hot electrons that 
produce both synchrotron emission in the microwaves and Inverse-Compton emission of the microwave photons in the Gamma-rays. It has been proposed that this spherical 
distribution of a population of hot electrons has been created by some large episode of energy injection in the Galactic center over the past 10 Myrs, by either a period of 
activity of our super-massive black hole or a dramatic nuclear starburst (Dobler et al., 2010).  
A similar bipolar shell of similar sizes had also been previously detected in H$\alpha$ and continuum infrared (IR) emission, and attributed to 
dust entrained in a large-scale, bipolar wind powered by either a central starburst (Bland-Hawthorn \& Cohen, 2003) or Seyfert-like activity of our super-massive black hole (Bland-Hawthorn et al., 2013). 

\noindent
An association between the absorbed thermal X-ray emission seen towards the Galactic center and the Fermi Bubbles has also recently been proposed by Kataoka and 
collaborators (2015), who detect a spatially resolved modulation of the EM across the edges of the Fermi Bubbles and model it by assuming a radial 
density profile consisting of an empty innermost 3-kpc radius sphere followed by a 2-kpc thick shell of constant density at the edges of the bubbles, and a $\beta$-like 
density profile model outwards, with the same parameters found by Miller \& Bregman (2013). 
This general scenario is consistent with our findings, and our best-fitting models A and B can 
reproduce not only the average observed values of EM towards the Galactic center, but also its modulation across the bubbles. However our analysis of the LGL absorbers 
proves that the Galaxy's plane is also filled with hot and dense gas, which must therefore be properly accounted for in the modeling of the X-ray emission from the central 
regions of the Galaxy. 

\noindent
Finally, the Fermi Bubbles have also been detected recently in moderately ionized metal absorption, towards a single line of sight passing through the bubble and showing 
two velocity peaks, one blue-shifted and one red-shifted by few hundreds of km s$^{-1}$, compared to the rest-frame position of the transitions (Fox et al., 2015). 
These two peaks in velocity have been interpreted as due absorption by the near- and far-side of the bubble, as seen from our position, and would indicate an expansion 
velocity of the Bubble of about 1000 km s$^{-1}$ (Fox et al., 2015). 
If this interpretation is correct, the Fermi Bubbles would not only contain hot Compton-scattering electron responsible for the observed 
Gamma-ray emission, but also much cooler gas producing low-ionization metal absorption. Our findings shows that the same (or a similar) structure is also present in 
hotter, million degree, gas, traced by OVII absorption. 

\section{Acknowledgements}
F.N. thanks P. Ranalli for the extremely useful and fruitful discussion on {\em frequentist} versus bayesian statistical approaches. 
F.N. and F.S. acknowledge support from the INAF-PRIN grant 1.05.01.98.10. 
Y.K. acknowledges support from the grant PAIIPIT IN104215. 


\begin{table} 
\footnotesize
\begin{center}
\caption{\bf Separate HGL and LGL Fits}
\vspace{0.4truecm}
\begin{tabular}{|ccccccccc|}
\hline
Model & Model & $n_0$& $R_c$ or $\rho_c$ & $h_c$ & $R_s$ & Halo Size & Mass & $chi^2(dof)$ \\
Name & Type & ($10^{-2}$ cm$^{-3}$) & (kpc) & (kpc) & (kpc) & (kpc) & ($10^9$ M$_{\odot}$ & \\
\hline
M1 & Exp-SS & $4.9_{-0.4}^{+1.1}$ & $3.1_{-0.2}^{+0.3}$ & N/A & $5.4_{-0.4}^{+0.6}$ & $> 46$ & $3.3_{-1.4}^{+4.1}$ & 10.9(15) \\
M2 & Exp-CS & $52_{-15}^{+5}$ & $2.4_{-0.1}^{+0.3}$ & $0.16_{0.03}^{+0.04}$ & N/A & N/A & $0.14_{-0.06}^{+0.11}$ & 12.8(10) \\
\hline
\end{tabular}
\end{center}
\end{table} 

\begin{table} 
\footnotesize
\begin{center}
\caption{\bf Simultaneous LGL+HGL Fits}
\vspace{0.4truecm}
\begin{tabular}{|ccccccccc|}
\hline
Model & Model & $n_0$& $R_c$ & $\beta$ & $R_s$ & Halo Size & Mass & $chi^2(dof)$ \\
Name & Type & ($10^{-2}$ cm$^{-3}$) & (kpc) & & (kpc) & (kpc) & ($10^{11}$ M$_{\odot}$ & \\
\hline
A & $\beta$-SS & $2.7_{-0.3}^{+0.3}$ & $2.1_{-0.2}^{+0.3}$ & $0.62_{-0.04}^{+0.04}$ & $5.6_{-0.6}^{+0.6}$ & $> 64$ & $0.2_{-0.1}^{+0.3}$ & 28.7(27) \\
B & M2 + $\beta$-SS & $1.3_{-0.2}^{+0.3}$ & $0.7_{-0.1}^{+0.2}$ & $0.33_{0.03}^{+0.03}$ & $6.7_{-1/8}^{+0.9}$ & $> 193$ & $1.3_{-0.7}^{+2.1}$ & 30.9(27) \\
M3 & M2 + $\beta$-SS & $0.8_{-0.1}^{+0.1}$ & $1.7_{-0.2}^{+0.2}$ & 0.33 (frozen) & 0 (frozen) & $> 250$ & $2.0_{-0.5}^{+0.4}$ & 34.9(29) \\
M4 & $\beta$-SS & $5.8_{-0.9}^{+0.7}$ & $2.5_{-0.2}^{+0.2}$ & 0.62 (frozen) & 0 (frozen) & $> 94$ & $0.5_{-0.1}^{+0.2}$ & 45.6(29) \\
\hline
\end{tabular}
\end{center}
\end{table} 

\begin{figure}[h]
\begin{center}
\hbox{
\hspace{1.0cm}
\psfig{figure=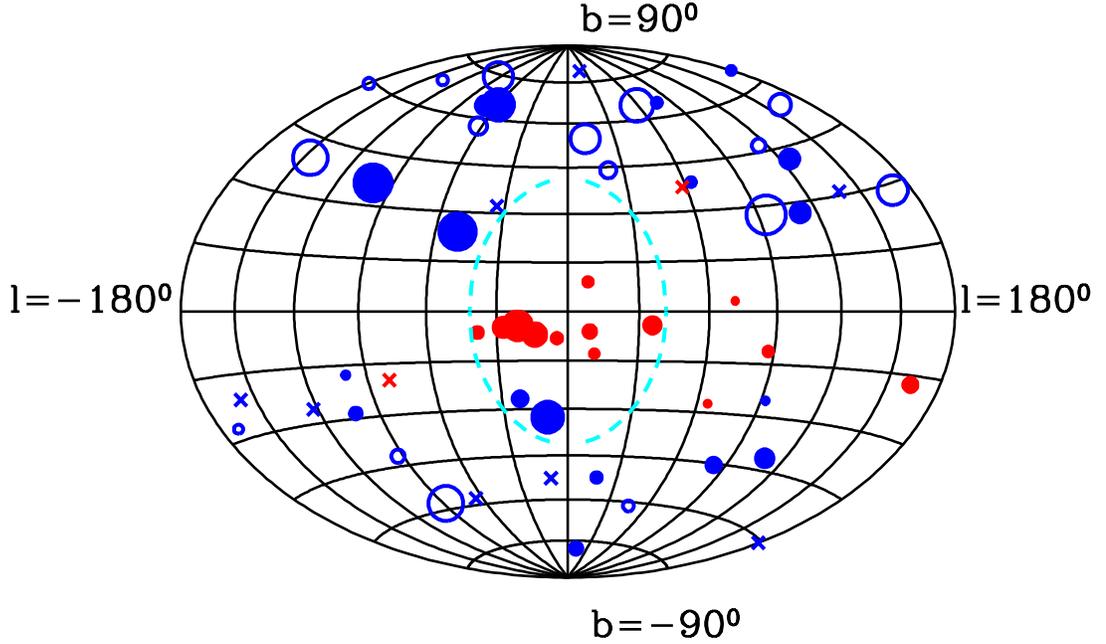,height=15.0cm,width=15.0cm,angle=0}
}
\end{center}
\vspace{-1.5cm} 
\caption{\footnotesize Aitoff projection of our LGL (red) and HGL (blue) absorbers. Filled circles are the 18 HGL plus 13 LGL measurements we use in our work. Empty circles 
are detections at significances lower than 2σ, while stars are upper limits or detections hampered by the presence of an instrumental feature. The size of the circles is 
proportional to the measured EW. Clearly most of the detections are concentrated towards the general direction of the center, with only 3 non-detections at 
$300^0 < l < 60^0$, and all three at $|b|>35^0$.  The LGL targets are mostly concentrated in the Galaxy's center, and this is why they are so powerful in constraining the 
central density and the disk profile (model M2 in Table 1). The dashed cyan line delimits the central cavity with radius $R_s = 5.6$ kpc, defined by our best-fitting model A.}
\label{fig1}
\end{figure}

\begin{figure}[h]
\begin{center}
\hbox{
\hspace{1.0cm}
\psfig{figure=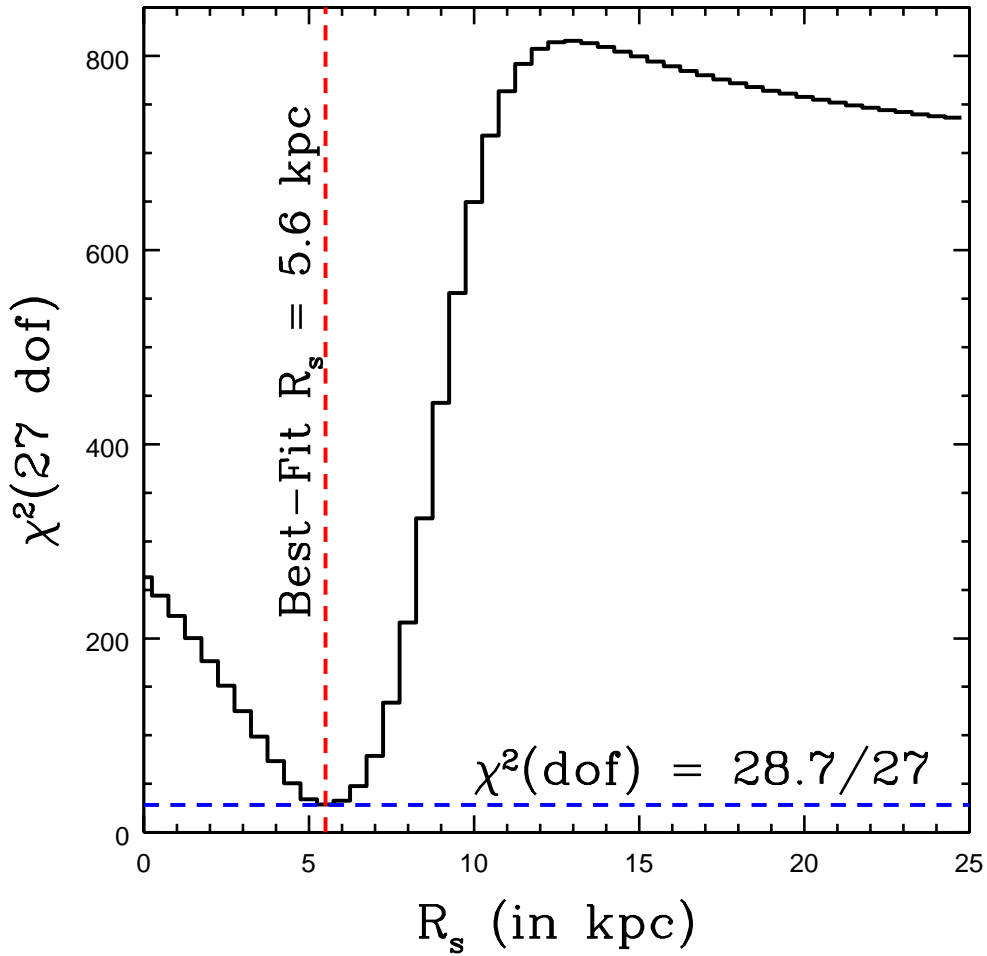,height=15.0cm,width=15.0cm,angle=0}
}
\end{center}
\vspace{-1.5cm} 
\caption{\footnotesize $\chi^2$ versus offset radius $R_s$, for our best-fitting model A. $\chi^2$s are for 1 interesting parameter, i.e. for normalizing density $n_0$, core radius 
$R_c$ and $\beta$-index, frozen to their best-fitting values (Table 2), versus the fourth parameter of the model, the offset radius $R_s$.  The sharp minimum at 5.6 kpc shows 
that values of $R_s$ lower than $R_s \simeq 3$ kpc ($\chi^2/dof \simeq 100/27$) and higher than $R_s \simeq 8$ kpc ($\chi^2/dof \simeq 150/27$) are clearly ruled out at 
high confidence}
\label{fig2}
\end{figure}

\begin{figure}[h]
\begin{center}
\hbox{
\hspace{1.0cm}
\psfig{figure=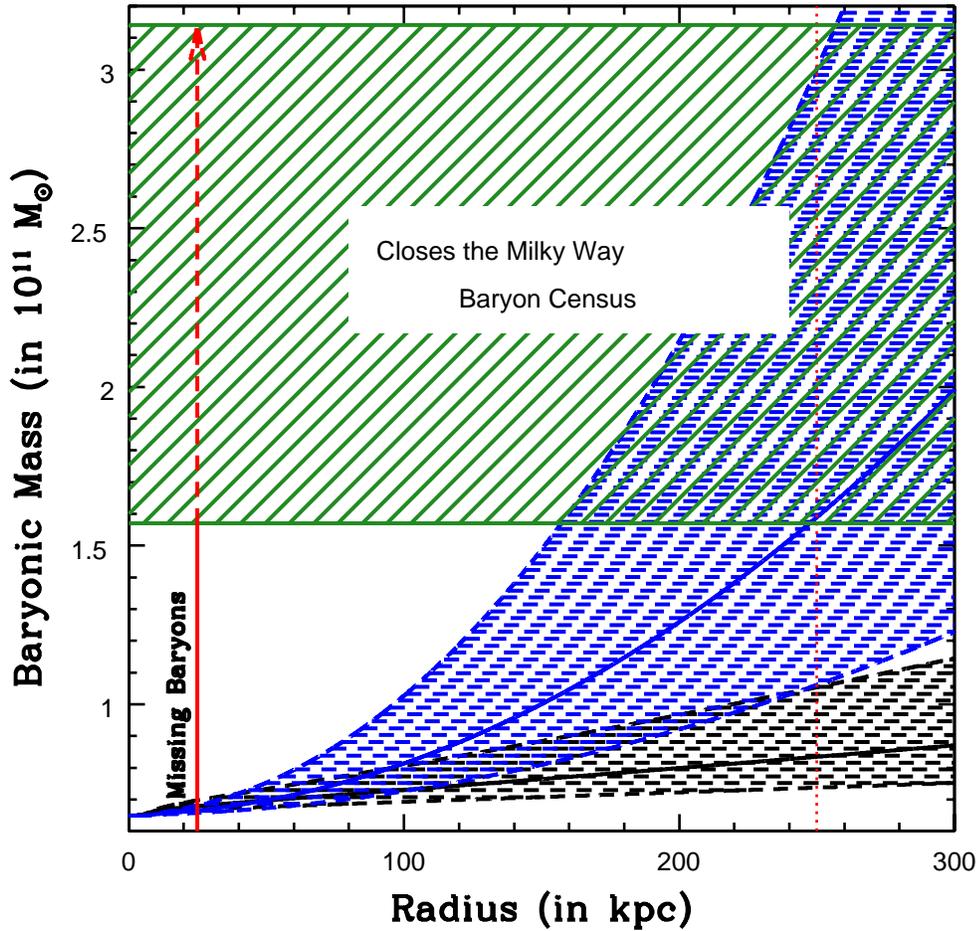,height=15.0cm,width=15.0cm,angle=0}
}
\end{center}
\vspace{-1.5cm} 
\caption{\footnotesize Total baryonic mass of the Milky Way as a function of the distance from the center. Blue and dark shaded regions show the 90\% confidence interval of 
Milky Way total  (i.e. visible + hot gas) baryon mass as derived from our best-fitting models A (black) and B (blue) and integrated over the Galactic radius from R=0 up to 1.2 
virial radii (the dashed red vertical line indicates 1 Galaxy's virial radius). The green shaded region highlights the range of baryon mass needed to close the Galaxy's baryon 
census. At 1.2 virial radii, model A predicts a mass that is still ~25\% less than the minimum needed to close the baryon census, while the flatter density profile model B 
accounts for all the needed baryon mass already at R=0.6 virial radii. Likely the actual solution lies in between models A and B.}
\label{fig3}
\end{figure}

\end{document}